\documentclass[twoside]{photon2007}
\usepackage[latin1]{inputenc}
\usepackage[dvips]{graphicx,epsfig,color}
\usepackage{wrapfig,rotating}
\usepackage{amssymb,amsmath,array}
\usepackage{amsmath,bm}

\begin{document}

\begin{figure}[htb]
\includegraphics{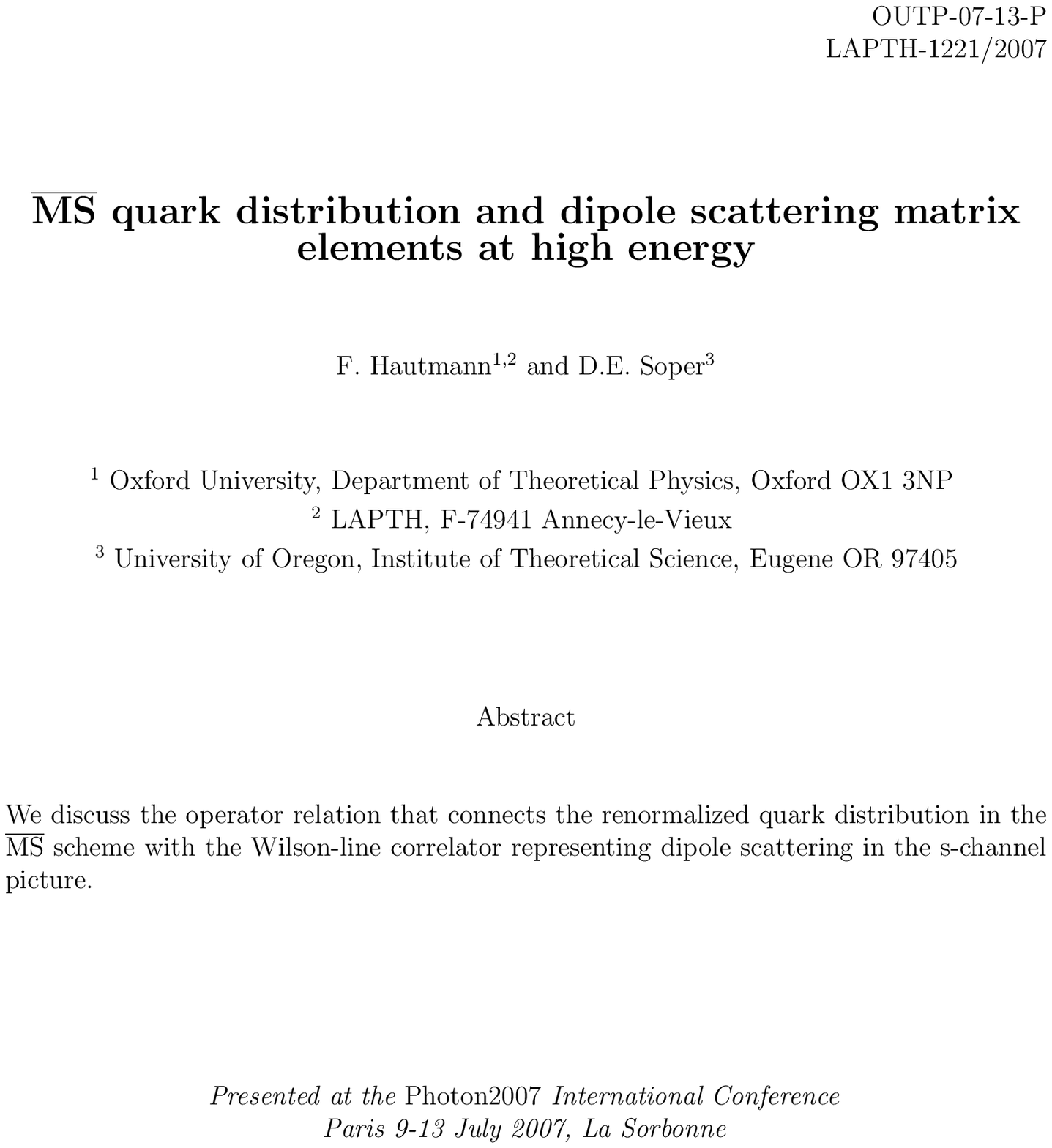}
\end{figure}

\title{
$\overline{\rm MS}$ quark distribution and dipole 
scattering matrix elements at high energy } 
\author{F.~Hautmann$^{1,2}$ and D.E.~Soper$^3$
\vspace{.3cm}\\
1-  Oxford University, Department of Theoretical Physics, Oxford OX1 3NP
\vspace{.1cm}\\
2-  LAPTH, F-74941 Annecy-le-Vieux   
\vspace{.1cm}\\
3-  University of Oregon, Institute of Theoretical Science, Eugene OR 97405 
}

\maketitle

\begin{abstract}
We discuss the operator relation 
that connects 
the renormalized quark distribution 
 in the $\overline{\rm MS}$ scheme 
with the Wilson-line correlator 
representing  dipole scattering in the 
s-channel picture.   
\end{abstract}

\vskip 1 cm

\noindent 
Parton picture and s-channel picture provide 
dual descriptions of high-energy hadron scattering. 
Based on the analysis~\cite{hs07},  
we here present the relation between 
the $\overline{\rm MS}$ parton distribution for quarks 
and the dipole scattering matrix elements  
that enter in the  s-channel approach to hard collisions. 
An introduction to the 
s-channel physical picture can be found 
in~\cite{bj90s}. For motivation based on  
QCD at high parton density,   
see~\cite{ericemue}-\cite{ianmue07} and references therein.

 Consider the 
 quark distribution function,  
defined as hadronic matrix element of a certain product 
of operators~\cite{cs82},    
\begin{eqnarray}
\label{eq:fqdef}
&& f_{q}(x,\mu) = 
{ 1 \over 4 \pi} 
\int\! d y^- e^{i xP^+y^-}
\\
&& \times   \langle P |
\bar\psi(0) Q(0) \gamma^+ 
Q^\dagger(y^-)\psi(0,y^-,{\bm 0})
|P\rangle_c  
\nonumber
\end{eqnarray}
where $\psi$ is the quark field, 
 $Q$ is the gauge link, and 
 the subscript $c$ is the instruction to take 
connected graphs.  
The matrix element (\ref{eq:fqdef}) 
can be rewritten 
as the real 
part of a forward scattering amplitude~\cite{hs07},      
 in which we  think of the operator $Q^\dagger \psi$  
as creating an antiquark 
plus an eikonal line  in the minus direction, starting  
at distance $y^-$ from the position of the target. 
The operator product in Eq.~(\ref{eq:fqdef}) has ultraviolet 
divergences and requires renormalization. 

\begin{figure}[htb]
\vspace{45mm}
\includegraphics{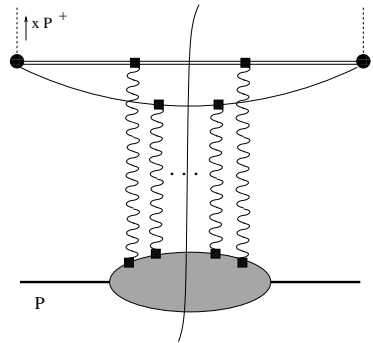}
\caption{Parton distribution function for quarks 
in the s-channel picture.} 
\label{fig:sec2}
\end{figure}

Next suppose  that the momentum fraction $x$ is very small.
 This means that the typical distance $y^-$ 
 from the target to 
where the antiquark and the eikonal line  
are created is large, 
of order $1/(x P^+)$ (Fig.~\ref{fig:sec2}). 
In the proton rest frame this is 
 far outside the proton~\cite{bj90s,ericemue}. 
We describe 
the  evolution of the parton system  
in the s-channel  using the 
 hamiltonian techniques~\cite{hks}. 
This allows one to express 
the evolution operator 
 in the high-energy approximation 
as an expansion in Wilson-line matrix elements, 
the leading term of which is the dipole term 
\begin{eqnarray}
\label{xidef}
\Xi( {\bm \Delta}, {\bm b}) = 
\int [ d P^\prime ] \   
\langle P'|\frac{1}{N_c} \ {} {\rm Tr}  \{ 1
\nonumber\\
 - F^{\dagger}( {\bm b} + {\bm \Delta}/2)\,
F({\bm b}-{\bm \Delta}/2)
\} |P \rangle \hspace*{0.2 cm} ,  
\end{eqnarray}       
\noindent where, in the  
 notation of~\cite{hs00},   $F$ is the non-abelian eikonal phase, 
${\bm \Delta}$ is the 
transverse separation between the eikonals, 
and ${\bm b}$ is the impact parameter.

In this s-channel       
  representation,   the quark distribution (\ref{eq:fqdef}) 
is given by  the coordinate-space convolution 
\begin{eqnarray}
\label{convfq}
x f_q ( x , \mu) &=& \int  {d{\bm b}} \ {d{\bm \Delta}} \ 
u (\mu ,  {\bm \Delta} ) \ \Xi( {\bm \Delta}, {\bm b})  
\nonumber\\
&-& UV \;\; . 
\end{eqnarray}
 In~\cite{hs07}   the explicit result  is given for the 
function $u (\mu ,  {\bm \Delta} )$ at one loop  
in dimensional regularization and for the counterterm $- UV$ of   
  $\overline {\rm MS}$  renormalization. 

The one-loop 
 $\overline {\rm MS}$ result  can be recast 
in a physically  more transparent  form 
in terms of  a 
cut-off  on the ${ \Delta}$ integration region. This is 
 defined by 
${ \Delta} \mu > a$, with $a$ a parameter of order 1. 
To do this,  the 
main 
point is to use the $- UV$ 
counterterm to 
cancel the  ultraviolet divergence,  and 
to determine $a$ so  as to cancel 
the finite remainder from the 
small-${ \Delta}$ integration region~\cite{hs07}. 
This yields 
\begin{eqnarray}
\label{twopiece}
&& xf_{q/p}(x,\mu) 
\\
&=& 
\frac{N_c}{3 \pi^4}\
\!\int\! {d{\bm b}} \ d{\bm \Delta} \
\theta(\Delta^2\mu^2 > a^2)\
\frac{\Xi({\bm \Delta}, {\bm b})}{\Delta^4} 
\nonumber\\ 
&+&
\frac{N_c}{3 \pi^4}\
\!\int\!  d{\bm \Delta} \ 
\theta(\Delta^2\mu^2 < a^2)\
\frac{R({\bm \Delta})}{\Delta^4}  , 
\nonumber
\end{eqnarray}   
where  $R$ is the ${\cal O} ({ \Delta}^4)$ 
remainder from  the  expansion of the ${\bm b}$-integral of 
$\Xi$ near 
${ \Delta} = 0 $, and $a$ is a calculated number that 
accomplishes  $\overline {\rm MS}$ renormalization,  
\begin{equation}
\label{eq:aresult}
a = 2 e^{1/6 - \gamma} \approx 1.32657 \;\;  , 
\end{equation}
with $\gamma$ the Euler constant. As long as $\mu$ is sufficiently 
large compared to the inverse hadron radius, the last term in the 
right hand side of Eq.~(\ref{twopiece}) is suppressed by powers 
of $\mu$, and one can write  
\begin{eqnarray}
\label{eq:renormalizedfq}
&& xf_{q/p}(x,\mu) 
\\
&=&  
\frac{N_c}{3 \pi^4}\
\!\int\! {d{\bm b}} \ { {d{\bm \Delta}} \over  { {\Delta^4}}  }  \  
\theta(\Delta^2\mu^2 > a^2) \
\Xi({\bm \Delta} , {\bm b} ) 
\nonumber   
\end{eqnarray}

Eq.~(\ref{eq:renormalizedfq}) is 
a remarkably simple formula.  
 The power behavior $1 / \Delta^4$ is  
perturbative, and 
$a$ is a renormalization scheme dependent coefficient, 
 Eq.~(\ref{eq:aresult}).  
The Wilson-line  matrix element 
 $\Xi({\bm \Delta}, {\bm b})$ 
receives contribution from both long  distances and 
short distances. 
At large ${ \Delta}$ (saturation region)~\cite{golrev}, 
 it has to be modeled or fit to 
data, while at small ${\Delta}$ it 
  can be treated by a short distance expansion.  
 
In particular, Ref.~\cite{hs07} uses  
one-loop renormalization-group evolution 
 equations  
to relate $\Xi({\bm \Delta}, {\bm b})$  at 
small ${\Delta}$ to a 
well-defined integral of the 
 gluon distribution function. 
 This  relation represents the  
 expression (at one loop) of 
color transparency. 
 It will be of interest  to study this relation for  the 
currently accessible $x$, jointly 
with  Eq.~(\ref{eq:renormalizedfq}), 
as a  way to probe   color transparency  
at the level  of  
nucleon's parton distribution functions~\cite{hs07,hs00}.

The method described above can also be applied to 
quantities that have a more direct physical interpretation, e.g.  
structure functions. In particular,  
this is used in~\cite{plb06} to identify the power-like 
contributions  
to the $Q^2$-derivative of $F_2$ 
 that come   from  multiple gluon 
scatterings. 
These  are  enhanced as  $x \to 0$, 
consistently with observations of approximate geometric 
scaling in low-$x$ data~\cite{geoscal}.      
Since the $F_2$  data used at present to extract 
the gluon density for $x < 10^{-2}$ do not have very high 
$Q^2$, these power corrections 
 can be relevant to the estimate of the 
theoretical accuracy on  small-$x$ pdf's for the LHC.  
If so, their 
effect  is to be taken into account along with 
  that~\cite{xlog} of 
higher-order $\ln x$ corrections.

We conclude 
by observing that   
the main applications of the results discussed in this 
article 
are to processes dominated by the sea quark distribution 
at  small $x$. Besides  collider applications,  
 these include the estimate of cross sections 
for high-energy cosmic neutrinos. 
Recent work in~\cite{berger}  
 proposes incorporating   
unitarity corrections to 
DGLAP predictions~\cite{sarkars,reno}  for 
 high-energy $\nu$ scattering,  
 based on a parton-model 
 fit to   $F_2$ data that enforces  the 
Froissart bound. 
Note that, if  parton 
degrees of freedom are  relevant  
at cosmic shower energies,  we expect 
neutrino interactions to be  
 dominated by the s-channel dynamics 
 of the sea quark, as represented in  
Eq.~(\ref{eq:renormalizedfq}) and Fig.~\ref{fig:sec2}.

\vskip 0.6 cm 

\noindent {\bf Acknowledgments.} F.H. 
thanks the organizers 
for their  hospitality, and  the 
  pleasant atmosphere at the 
conference. 

\vskip 0.7 cm

\begin{footnotesize}

\end{footnotesize}

\end{document}